\documentclass[10pt,a4paper]{article}

\usepackage{graphicx}
\usepackage{amssymb,amsmath}
\usepackage{a4wide}

\newtheorem{thm}{Theorem}
\begin{document}

\title{The Capacity of the Single Source Multiple Relay Single Destination Mesh Network}

\author{Lawrence Ong and Mehul Motani\\
Department of Electrical and Computer Engineering\\
National University of Singapore\\
Email: \{lawrence.ong, motani\}@nus.edu.sg}
\date{}

\maketitle

%ISIT standard: not more than 250 words
\begin{abstract}
In this paper, we derive the capacity of a special class of mesh networks. A mesh network is defined as a heterogeneous wireless network in which the transmission among power limited nodes is assisted by powerful relays, which use the same wireless medium. We find the capacity of the mesh network when there is one source, one destination, and multiple relays. We call this channel the single source multiple relay single destination (SSMRSD) mesh network. Our approach is as follows. We first look at an upper bound on the information theoretic capacity of these networks in the Gaussian setting. We then show that the bound is achievable asymptotically using the compress-forward strategy for the multiple relay channel. Theoretically, the results indicate the value of cooperation and the utility of carefully deployed relays in wireless ad-hoc and sensor networks. The capacity characterization quantifies how the relays can be used to either conserve node energy or to increase transmission rate.
\end{abstract}

\section{Introduction}
Wireless networks have been finding more applications and capturing much research attention in recent years. The prevalence of mobile devices makes the wireless network an attractive solution for home and enterprise users.  Unfortunately, the analysis of these multi-terminal networks is difficult. To date, the capacity of even the simple three-node channel~\cite{meulen71} is not known, except for special cases, for example, the multiple access channel \cite{liao72}\cite{ahlswede74}, the degraded relay channel \cite{covergamal79}, the degraded broadcast channel \cite{bergmans73}. However, this did not hinder research in channels with more nodes.

A natural extension of the single source single destination three-node channel to the multiple node channel is the multiple relay channel \cite{xiekumar03}\cite{kramergastpar04}\cite{ongmotani05a}\cite{ongmotani05b}\cite{chongmotani05b}. The multiple relay channel captures the scenario where the transmission from the source to the destination is aided by relay nodes, which themselves have no data to send. One can also treat this as an excerpt of a general multi-terminal network, where we consider just one of the source-destination pairs. The capacity of the multiple relay channel has not been found except for the degraded multiple relay channel \cite{xiekumar03}. In this paper, we investigate the capacity of another class of multiple relay channels -- the single source multiple relay single destination (SSMRSD) \emph{mesh network} .

\begin{figure}[t]
\centering
\includegraphics[width=9cm]{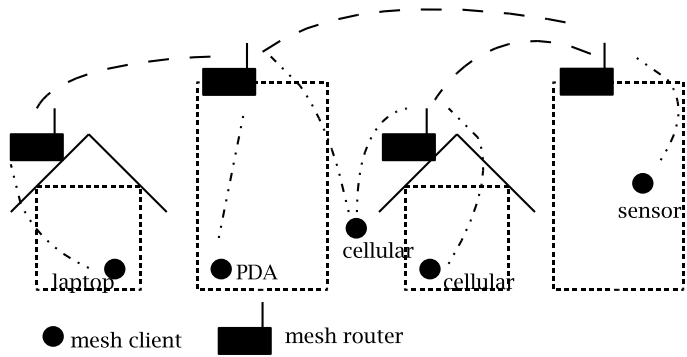}
\caption{A mesh network.}  \label{fig:mesh}
\end{figure}

The mesh network (see \cite{akyildiz05} and the references therein) is a multi-terminal channel with powerful relays. One practical setup of the mesh network is depicted in Fig.~\ref{fig:mesh}. Mesh routers (powerful relay nodes connected to power supplies) are installed on top of houses and buildings. These routers communicate with various mesh clients (source nodes with average power constraint or destination nodes) in their proximity and connect to other mesh routers. The area between buildings are covered and any two mesh clients can send data to each other which might not have been possible without the mesh routers. The routers are able to help the source to send data at a higher rate to the destination. We note that even though the mesh routers are not bounded by restricted battery lifetime as they are connected to the power line, their transmit power is often restricted by regulations. However, the study of mesh network is still interesting as it gives insights on how nodes should cooperate when the relays can transmit at higher power (which might not be infinity) compared to the sources.

In this paper, we consider the SSMRSD mesh network, in which there is only one source and one destination but any number of relays. We note that the SSMRSD mesh network is not a degraded multiple relay channel \cite[Theorem 3.2]{xiekumar03}. The capacity of these channels has not been found.

Gupta and Kumar \cite{guptakumar00} considered a general wireless network model, in which every node has data to send to a random destination.  In this scenario, they determined the scaling behavior of the transport capacity of the network with respect to the number of nodes in the network.  The mesh network differs from their model as mesh routers in the mesh network do not generate data.

In \cite{junsichitiu03} and \cite{jun02}, the authors found the practical ``capacity'' of the mesh network with the following assumptions:
\begin{itemize}
	\item All nodes send data to a common gateway.
	\item Each node is given a fair amount of bandwidth.
	\item The physical layer and the MAC layer is assumed to follow the 802.11 standard.
	\item A proper transmission scheduling scheme is used to avoid node interference.
\end{itemize}
Our work attempts to find the capacity (in an information theoretic sense) of the mesh network without any constraints on the physical and the MAC layers.

Our approach is as follows. First we study an upper bound on the capacity of the SSMRSD mesh network, which is derived from the max-flow min-cut theorem. Then we study an achievable rate of the compress-forward non-constructive coding strategy on the multiple relay channel. The technique was first introduced in \cite{covergamal79} for the single relay channel and later extended to the multiple relay channel in \cite{kramergastpar04}, where it is called the compress-and-forward strategy. We show that when the transmit powers of the relays increase, the compress-forward technique approaches the capacity upper bound asymptotically.

The rest of the paper is organized as follows. Section~\ref{mesh_channel} introduces the channel models and definitions. In Section~\ref{mesh_theories}, we establish several useful theorems that we will need in later sections. In Section~\ref{mesh_upper_bound}, we investigate an upper bound on the capacity of the SSMRSD mesh network. This is followed by studying achievable rates on the multiple relay channel in Section~\ref{mehs_mrc}. By looking at the special channel, i.e., when the relays have no power constraint, we show in Section~\ref{mesh_capacity_ssmr} that the achievable rate of the Gaussian SSMRSD mesh network approaches the capacity of the channel asymptotically. Section~\ref{sec:mesh_conclusion} concludes the paper.

\section{Channel Model} \label{mesh_channel}
\begin{figure}[ht]
\centering
\includegraphics[width=5.5cm]{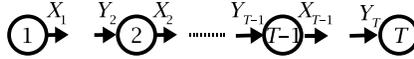}
\caption{The multiple relay channel.}  \label{fig:ssmr}
\end{figure}

Fig.~\ref{fig:ssmr} depicts the  multiple relay channel. The  multiple relay channel can be
completely described by the channel distribution $p^*(y_2, y_3, \dotsc, y_T | x_1, x_2, \dotsc, x_{T-1})$ on $\mathcal{Y}_2 \times \mathcal{Y}_3 \times \dotsm \times \mathcal{Y}_T$, for each $(x_1, x_2, \dotsc, x_{T-1}) \in \mathcal{X}_1 \times \mathcal{X}_2 \times \dotsm \times \mathcal{X}_{T-1}$.  In this paper, we only consider memoryless channels. Node 1 is the source node and node $T$ is the destination node.  Nodes 2 to $T-1$ are purely
relay nodes. Message $W$ is generated at node 1 and is to be transferred to the
sink at node $T$.
We follow the definitions of capacity, achievable rate ($R_W$) used in \cite[Section III.A]{kramergastpar04}.

In a Gaussian multiple relay channel, node $j$ receives
\begin{equation}
Y_j = \sum_{\substack{i=1, \dotsc, T-1 \\ i \neq j}} \sqrt{\lambda_{ij}}X_i + Z_j, \quad j=2, \dotsc, T,
\end{equation}
where $X_i$, input to the channel form node $i$, is a random variable with power constraint $E[X_i^2] \leq P_i$. $Y_j$ is the received signal at node $j$. $Z_j$, the receiver noise at node $t$, is an independent zero mean Gaussian random variable with variance $N_j$.  $\lambda_{ij} = \kappa d_{ij}^{-\eta}$ is the path loss function. $d_{ij}$ is the distance between node
$i$ and node $j$, $\eta$ is the path loss exponent, and
$\eta \geq 2$ with equality for free space transmission. $\kappa$ is a
positive constant.

The $T$-node Gaussian SSMRSD mesh network is defined as the $T$-node Gaussian multiple relay channel where $\frac{P_i}{P_1} \gg 1$ for all $i \in \mathcal{R}$. We define $\mathcal{R} \triangleq \{ 2, 3, \dotsc, T-1\}$ as the set of all relay nodes. We use the notation $X_{\{1, \dotsc, m\}} \triangleq (X_1, \dotsc, X_m)$.

\section{A Cut-Set Bound is Attained by Independent Gaussian Inputs} \label{mesh_theories}
In this section, we establish a useful theorem which we will need in later sections. In brief, we consider the Gaussian relay channel where the relay(s) and the destination can cooperate. The following theorems establish that the optimal input distribution to maximize the mutual information between the source node, and the relays plus the destination is such that the the source and the relays send independent Gaussian inputs.

We consider a $T$-node multiple relay channel where nodes $1,\dotsc,T-1$ send $X_1, \dotsc, X_{T-1}$ into the channel respectively. The channel inputs are subject to power constraints $E[X_i^2] \leq P_i$ for $i=1,\dotsc,T-1$. Without loss of generality, nodes $2, \dotsc, T$ receive the following signals from the channel.
\begin{equation}
Y_j = \sum_{i \in \{1\} \cup \mathcal{R} \setminus \{j\}} X_i + Z_j,
\end{equation}
where $Z_j \sim \mathcal{N}(0,N_j)$, $j=2,3, \dotsc, T$ are independent Gaussian noise. Here, we ignore the path loss component for simplicity. The results hold for channels with the path loss component.

\begin{thm}\label{thm:mrc_gen_indep_gauss}
Consider a $T$-node Gaussian multiple relay channel. A sufficient condition on the input distribution that achieves
\begin{equation}\label{eq:mutual_mrc_gen_indep_gauss}
\max_{p(x_1,x_2, \dotsc, x_{T-1})} I(X_1;Y_\mathcal{R},Y_T|X_\mathcal{R})
\end{equation}
is that the inputs are Gaussian and $X_1$ is independent of $X_\mathcal{R}$. It follows that independent Gaussian inputs $X_1, \dotsc, X_{T-1}$ also achieve \eqref{eq:mutual_mrc_gen_indep_gauss}.
\end{thm}

\emph{Proof:} First, we consider the case $T=3$, which means there is one relay. We want to show that 
\begin{equation}
\max_{p(x_1,x_2)} I(X_1;Y_2,Y_3|X_2)
\end{equation}
is achieved when $X_1$ and $X_2$ are independent Gaussian inputs.

From \cite[Proposition 2]{kramergastpar04}, we know the optimal input distribution is Gaussian. We let
\begin{equation}
X_1 = \alpha X_2 + W,
\end{equation}
where $W$ and $X_2$ are independent Gaussian random variables, such that $E[W^2]=P_W$ and $P_1 = \alpha^2P_2 + P_W$.

Now,
\begin{equation}
H(Y_2,Y_3|X_1,X_2) = \frac{1}{2} \log (2\pi e)^2N_2N_3,
\end{equation}
and
\begin{subequations}
\begin{align}
H(Y_2,Y_3|X_2) & = \frac{1}{2} \log (2\pi e)^2 
\begin{vmatrix}
P_W + N_2 & P_W \\
P_W  & P_W + N_3
\end{vmatrix} \\
& = \frac{1}{2} \log (2\pi e)^2 (P_WN_2 + P_WN_3 + N_2N_3).
\end{align}
\end{subequations}

Hence,
\begin{subequations}
\begin{align}
I(X_1;Y_2,Y_3|X_2) & = H(Y_2,Y_3|,X_2) - H(Y_2,Y_3|X_1,X_2) \\
& = \frac{1}{2} \log \left[ 1 + \frac{P_1 - \alpha^2 P_2}{N_2} + \frac{P_1 - \alpha^2P_2}{N_3}  \right].
\end{align}
\end{subequations}
Setting $\alpha=0$ maximizes the mutual information. This completes the proof for $T=3$.

Now, we extend this result to $T=4$ or the two-relay channel. The generalization from the two-relay channel to the multiple-relay channel is straight forward. We need to show that a sufficient condition on the input distribution function to achieve
\begin{equation}\label{eq:mutual_mrc_indep_gauss}
\max_{p(x_1,x_2,x_3)} I(X_1;Y_2,Y_3,Y_4|X_2,X_3)
\end{equation}
is that $X_1$ and  $(X_2,X_3)$ are independent Gaussian inputs.

From \cite[Proposition 2]{kramergastpar04}, \eqref{eq:mutual_mrc_indep_gauss} is achieved by Gaussian inputs $X_1$, $X_2$, and $X_3$. From the single relay case $T=3$, we know that choosing $X_1$ to be independent of $(X_2,X_3)$ is optimal. Certainly, choosing independent $X_1$, $X_2$, and $X_3$ maximizes the mutual information term. This proves the case of $T=4$.

Now, we demonstrate that \eqref{eq:mutual_mrc_indep_gauss} can indeed be achieved with any correlation between $X_2$ and $X_3$, as long as $X_1$ is independent of $(X_2,X_3)$. We let $X_2 = \beta X_3 + W$, where $X_1$, $X_3$ and $W$ are independent Gaussian inputs. Here, $E[W^2]=P_W$ and $P_2 = \beta^2P_3 + P_W$.

Now,
\begin{equation}
H(Y_2,Y_3,Y_4|X_1,X_2,X_3) = \frac{1}{2} \log (2\pi e)^3 N_2N_3N_4.
\end{equation}
Also,
\begin{subequations}
\begin{align}
& H(Y_2,Y_3,Y_4|X_2,X_3) \nonumber \\
& = \frac{1}{2} \log (2\pi e)^3
\begin{vmatrix}
P_1+N_2 & P_1 & P_1 \\
P_1 & P_1+N_3 & P_1 \\
P_1 & P_1 & P_1+N_4
\end{vmatrix}\\
& = \frac{1}{2} \log (2\pi e)^3 \left[ P_1(N_2N_3+N_2N_4+N_3N_4) + N_2N_3N_4  \right].
\end{align}
\end{subequations}
Hence,
\begin{equation}
I(X_1;Y_2,Y_3,Y_4|X_2,X_3) = \frac{1}{2} \log \left[ 1 + P_1\left( \frac{1}{N_2} + \frac{1}{N_3} + \frac{1}{N_4}  \right) \right].
\end{equation}
We note that this is independent of $\beta$. This means that \eqref{eq:mutual_mrc_indep_gauss} can be achieved with any correlation between $X_2$ and $X_3$.

We can easily generalize this result to any $T > 4$ and hence obtain Theorem~\ref{thm:mrc_gen_indep_gauss}.

\section{An Upper Bound on the Capacity of the Multiple Terminal Network} \label{mesh_upper_bound}
\subsection{In the Multi-Terminal Network}
\begin{figure}[ht]
\centering
\includegraphics[width=3.6cm]{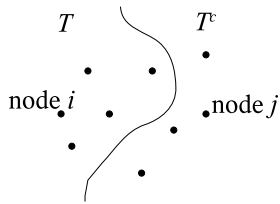}
\caption{A cut in the multi-terminal network.}  \label{fig:cut_multi_terminal}
\end{figure}

Consider a T-node multi-terminal network where node $i$ transmits $X_i$ and node $j$ receives $Y_j$. The channel is characterized by the channel transition probability $p(y_1, \dotsc, y_T | x_1, \dotsc, x_T)$. \cite[Theorem 14.10.1]{coverthomas91} states that if the rate from node $i$ to node $j$, $R_{ij}$, is achievable, then the following must be satisfied
\begin{equation}\label{eq:cut_rate}
\sum_{i \in \mathcal{T}, j \in \mathcal{T}^c} R_{ij} \leq \max_{p(x_1, \dotsc, x_T)}I(X_\mathcal{T} ; Y_{\mathcal{T}^c} | X_{\mathcal{T}^c} ),
\end{equation}
for some joint probability function $p(x_1, \dotsc, x_T)$ for all $\mathcal{T} \subset \{ 1, \dotsc, T\}$ where $i \in \mathcal{T}$ and $j \notin \mathcal{T}$. $\mathcal{T}^c$ is the complement of $\mathcal{T}$ in $\{1, \dotsc, T\}$.

We can interpret this theorem as follows. The achievable rate from node $i$ to node $j$ must be smaller than the rate of all possible cuts separating nodes $i$ and $j$. Fig.~\ref{fig:cut_multi_terminal} depicts a possible cut. We define the \emph{cut rate} for the cut separating $\mathcal{T}$ and $\mathcal{T}^c$ as the right side of \eqref{eq:cut_rate}. It is the maximum achievable rate from nodes in $\mathcal{T}$ to nodes in $\mathcal{T}^c$ when all nodes on the same side of the cut are allowed to cooperate.

\subsection{In the SSMRSD Mesh Network}
Consider a $T$-node Gaussian SSMRSD mesh network where
\begin{itemize}
	\item Node 1 is the source node with power constraint $E[X_1^2] \leq P_1$, which can only transmit.
	\item Node $T$ is the destination node, which can only receive signals from the network.
	\item Nodes 2 to $T-1$ are powerful relays with large power constraint, which can transmit and receive at the same time.
	\item None of the relays or the destination is close to the source.
\end{itemize}
We note that any cut rate with $1 \in \mathcal{T}$ and $T \in \mathcal{T}^c$ is an upper bound of the rate from the source to the destination. Since the relays have large power, if we include any relay node in set $\mathcal{T}$, the cut rate (defined as \eqref{eq:cut_rate}) is large. Hence the minimum cut rate occurs when the cut separates $\mathcal{T} = \{ 1 \}$ and $\mathcal{T}^c = \{2, \dotsc, T \}$. So the upper bound of the capacity of the SSMRSD reduces to
\begin{equation}\label{eq:cutset_mesh}
C_{\text{SSMRSDMesh}} \leq \max_{\substack{p(x_1, \dotsc, x_{T-1})\\E[X_1^2] \leq P_1}} I(X_1 ; Y_\mathcal{R},Y_T | X_\mathcal{R}),
\end{equation}
for some joint probability function $p(x_1, \dotsc, x_{T-1})$. From Theorem~\ref{thm:mrc_gen_indep_gauss}, independent Gaussian inputs maximize this upper bound in the Gaussian channel.

\section{Achievable Rates} \label{mehs_mrc}

\subsection{In the General Multiple Relay Channel}
In this section, we investigate achievable rates of the multiple relay channel using the compress-forward strategy. Using \cite[Theorem 3]{kramergastpar04} and setting $U_t = X_t$, $\forall t \in \mathcal{R}$, the following rate is achievable in the multiple relay channel using the compress-forward strategy,
\begin{equation}\label{eq:cf_rate_mesh}
R = I(X_1; \tilde{Y}_\mathcal{R} Y_T | X_\mathcal{R}),
\end{equation}
where
\begin{equation}\label{eq:cf_condition_mesh}
I(\tilde{Y}_\mathcal{S} ; Y_\mathcal{S} | X_\mathcal{R}, \tilde{Y}_{\mathcal{S}^c}, Y_T) \leq \sum_{m=1}^M I( X_{\mathcal{B}_m} ; Y_{r(m)} | X_{\mathbf{B}_m^c}),
\end{equation}
with the joint probability distribution function
\begin{equation}\label{eq:cf_distribution_mesh}
p(x_1) \left[ \prod_{t \in \mathcal{R}}p(x_t)p(\tilde{y}_t | x_{\mathcal{R}},y_t) \right] p^*(y_\mathcal{R}, y_T | x_1, x_\mathcal{R}),
\end{equation}
for all $\mathcal{S} \subseteq \mathcal{R}$, all partitions $\{\mathcal{B}_m \}_{m=1}^M$ of $\mathcal{S}$, and all $r(m) \in \{ 2, \dotsc, T\} \setminus \mathcal{B}_m$. $\mathcal{S}^c$ is the complement of $\mathcal{S}$ in $\mathcal{R}$ and $\mathcal{B}_m^c$ is the compliment of $\mathcal{B}_m$ in $\mathcal{R}$. $U$ is the part which is to be decoded by all relays. Setting $U_t = X_t$ means each relay decodes all other relays' codewords. We note that in the compress-forward strategy, all channel inputs $X_1, \dotsc, X_{T-1}$ are independent.

\subsection{In the Gaussian Multiple Relay Channel}

We consider the Gaussian multiple relay channel. By relaxing the power constraint on the relays, or nodes $t \in \mathcal{R}$, the multiple relay channel is equivalent to the SSMRSD mesh network.

Now, using the compress-forward strategy with $U_j = X_j$, the received signal of node $r(m)$ can be written as
\begin{subequations}
\begin{align}
Y_{r(m)} & = \sqrt{\lambda_{1r(m)}}X_1 + \sum_{\substack{i \in \mathcal{R} \\ i \neq r(m)}} \sqrt{\lambda_{ir(m)}}X_i + Z_{r(m)}\\
& = \sqrt{\lambda_{1r(m)}}X_1 + \sum_{\substack{i \in \mathcal{B}_m \\ i \neq r(m)}} \sqrt{\lambda_{ir(m)}}X_i \nonumber\\
& \quad + \sum_{\substack{i \in \mathcal{B}_m^c \\ i \neq r(m)}} \sqrt{\lambda_{ir(m)}}X_i  + Z_{r(m)}.
\end{align}
\end{subequations}

The term inside the summation on the right hand side of \eqref{eq:cf_condition_mesh} can be evaluated as
\begin{equation}
I( X_{\mathcal{B}_m} ; Y_{r(m)} | X_{\mathbf{B}_m^c}) = \frac{1}{2} \log \left[ 1 + \frac{\sum_{\substack{i \in \mathcal{B}_m \\ i \neq r(m)}} \lambda_{ir(m)}P_i}{\lambda_{1r(m)}P_1 + N_{r(m)}} \right].
\end{equation}
We note that all $X_i$ are independent, as seen from \eqref{eq:cf_distribution_mesh}.

Using the compress-forward strategy, the node $j$'s quantized received signal is
\begin{equation}
\tilde{Y}_j = Y_j + W_j = \sum_{\substack{i=1, \dotsc, T-1 \\ i \neq j}} \sqrt{\lambda_{ij}}X_i + Z_j + W_j,
\end{equation}
where $W_j \sim \mathcal{N}(0,Q_j)$ are independent quantization noise.

The left hand side of \eqref{eq:cf_condition_mesh} is
\begin{subequations}
\begin{align}
I(\tilde{Y}_\mathcal{S} ; Y_\mathcal{S} | X_\mathcal{R}, \tilde{Y}_{\mathcal{S}^c}, Y_T)
& \leq I(\tilde{Y}_\mathcal{S} ; Y_\mathcal{S} | X_\mathcal{R}) \label{reduce_entropy}\\
& = H(\tilde{Y}_\mathcal{S} | X_\mathcal{R}) - H(\tilde{Y}_\mathcal{S} | Y_\mathcal{S}, X_\mathcal{R})\label{eq:bound}
\end{align}
\end{subequations}
The first term in \eqref{eq:bound} is
\begin{equation}
H(\tilde{Y}_\mathcal{S} | X_\mathcal{R}) = \frac{1}{2} \log 2\pi e^D \Lambda(D),
\end{equation}
where $\Lambda(D)$ is defined as
\begin{equation}\label{eq:determinant}
\Lambda(D) =
\begin{vmatrix}
\lambda_{1s(1)}P_1 + N_{s(1)} + Q_{s(1)} & \dotsm & \sqrt{\lambda_{1s(1)}\lambda_{1s(D)}}P_1 \\
\vdots & \ddots & \vdots \\
\sqrt{\lambda_{1s(1)}\lambda_{1s(D)}}P_1 & \dotsc & \lambda_{1s(D)}P_1 + N_{s(D)} + Q_{s(D)}
\end{vmatrix},
\end{equation}
$s(i)$ are ordered elements in $\mathcal{S}$ and $D = \lvert \mathcal{S} \rvert$.

The second term in \eqref{eq:bound} is
\begin{equation}
H(\tilde{Y}_\mathcal{S} | Y_\mathcal{S}, X_\mathcal{R}) = \frac{1}{2} \log 2\pi e^D Q_{s(1)} \dotsm Q_{s(D)}.
\end{equation}

Now a sufficient condition for \eqref{eq:cf_condition_mesh} is 
\begin{equation}
I(\tilde{Y}_\mathcal{S} ; Y_\mathcal{S} | X_\mathcal{R}) \leq \sum_{m=1}^M I( X_{\mathcal{B}_m} ; Y_{r(m)} | X_{\mathbf{B}_m^c}),
\end{equation}
or in the Gaussian channel,
\begin{equation}
Q_{s(1)} \dotsm Q_{s(D)} \geq \frac{\Lambda(D)}{\prod_{m=1}^M \left[ 1 + \frac{\sum_{\substack{i \in \mathcal{B}_m \\ i \neq r(m)}} \lambda_{ir(m)}P_i}{\lambda_{1r(m)}P_1 + N_{r(m)}} \right]}.
\end{equation}

Hence, we have the following theorem on the $T$-node Gaussian multiple relay channel.

\begin{thm}\label{thm:cf_mesh_achievable}
Consider a memoryless $T$-node Gaussian multiple relay channel. Using independent Gaussian input $X_i$, $i=1, \dotsc, T-1$, with power constraints $E[X_i^2] \leq P_i$, the following rate is achievable
\begin{equation}\label{eq:rate_achievable_mesh_thm}
R = \max_{\substack{\text{independent Gaussian inputs}\\E[X_i^2] \leq P_i}}I(X_1; \tilde{Y}_\mathcal{R}, Y_T | X_\mathcal{R}),
\end{equation}
where $\tilde{Y}_j = Y_j + W_j$ and $W_j \sim \mathcal{N}(0,Q_j)$ are independent quantization noise.
The rate equation is subject to the constraints
\begin{equation}\label{eq:ssmrmesh_cap_condition}
Q_{s(1)} \dotsm Q_{s(D)} \geq \frac{\Lambda(D)}{\prod_{m=1}^M \left[ 1 + \frac{\sum_{\substack{i \in \mathcal{B}_m \\ i \neq r(m)}} \lambda_{ir(m)}P_i}{\lambda_{1r(m)}P_1 + N_{r(m)}} \right]},
\end{equation}
for all $\mathcal{S} \subseteq \mathcal{R}$, $\{s(1)... s(D)\}=\mathcal{S}$, all partitions $\{\mathcal{B}_m \}_{m=1}^M$ of $\mathcal{S}$, and all $r(m) \in \{ 2, \dotsc, T\} \setminus \mathcal{B}_m$. $\mathcal{R}$ is the set of all relays.
\end{thm}

For \eqref{eq:ssmrmesh_cap_condition} to hold, a sufficient condition is that $P_j, \forall j \in \mathcal{R}$, are large, $\Lambda(D)$ not too large, $\lambda_{1j}P_1, \forall j \in \mathcal{R}$ not too large. With these extra conditions, we have the capacity theorem in the next section.

We note that the achievability of \eqref{eq:rate_achievable_mesh_thm} makes use of the Markov lemma~\cite[Lemma 4.1]{berger77}, which requires strong typicality. Though strong typicality does not extend to continuous random variables, we can generalize the Markov lemma for Gaussian inputs and thus show that \eqref{eq:rate_achievable_mesh_thm} is achievable \cite{kramergastpar04}.

\section{The Capacity of the Gaussian SSMRSD Mesh Network} \label{mesh_capacity_ssmr}
By definition, mesh networks employ powerful relay nodes. Now, we study the case when the relay power constraint grows without bound and finite source transmit power, meaning,
\begin{subequations}
\begin{align}
P_1 & < \infty \\
P_i  & \rightarrow \infty, \quad \forall i \in \mathcal{R}.
\end{align}
\end{subequations}
While this may not be practical, it does allow us to characterize the capacity and to study how the rates scale with power.
We also assume that the relays and the destination are not near the source, meaning
\begin{equation}
\lambda_{1j} = K_i, \quad \forall j \in \mathcal{R} \cup \{T\},
\end{equation}
for some $K_i$ not large.
Under this condition, we can set
\begin{equation}
Q_i \rightarrow 0, \quad \forall i \in \mathcal{R},
\end{equation}
while \eqref{eq:ssmrmesh_cap_condition} can still be satisfied  for all $\mathcal{S} \subseteq \mathcal{R}$, all partitions $\{\mathcal{B}_m \}_{m=1}^M$ of $\mathcal{S}$, and all $r(m) \in \{ 2, \dotsc, T\} \setminus \mathcal{B}_m$.
When $Q_i \rightarrow 0$, the quantized received signals approach the received signals, that is
\begin{equation}
\tilde{Y}_i = Y_i + W_i \rightarrow Y_i^+,
\end{equation}
for all $\forall i \in \mathcal{R}$. 
The achievable rate in \eqref{eq:rate_achievable_mesh_thm} becomes
\begin{equation}\label{eq:rate_achievable_mesh}
R \rightarrow \max_{\substack{\text{independent Gaussian inputs}\\E[X_1^2] \leq P_1}} I(X_1; Y_\mathcal{R} Y_T | X_\mathcal{R}).
\end{equation}

We see that \eqref{eq:rate_achievable_mesh} has the same form as the capacity upper bound \eqref{eq:cutset_mesh} of the SSMRSD mesh network. The upper bound \eqref{eq:cutset_mesh} is maximized over all possible input distributions but the achievable rate \eqref{eq:rate_achievable_mesh} is achievable with independent Gaussian inputs. However, Theorem~\ref{thm:mrc_gen_indep_gauss} states that the cut-set upper bound is maximized by using independent Gaussian inputs. Hence, the compress-forward strategy approaches the cut-set upper bound of the SSMRSD mesh network asymptotically.  This is summarized in the following theorem.
\begin{thm}
The achievable rate of the compress-forward strategy approaches the capacity of the Gaussian SSMRSD mesh network (where no node is near the source), which is equivalent to the Gaussian multiple relay channel (where the relays and the destination are not near the source), asymptotically as the relay power grows relays. The capacity is given by
\begin{equation}\label{eq:ssmr_mesh_cap_thm}
C_{\text{SSMRSDMesh}} = \max_{\substack{\text{independent Gaussian inputs}\\E[X_1^2] \leq P_1}}I(X_1; Y_\mathcal{R} Y_T | X_\mathcal{R}).
\end{equation}
\end{thm}

We note that the capacity is achieved by driving $Q_i \rightarrow 0$ hence making $\tilde{Y}_i \rightarrow Y_i$. This can also be achieved by driving $\frac{\lambda_{ij}P_i}{\lambda_{1j}P_1 + N_j} \rightarrow \infty, \forall i,j \in \mathcal{R}$ and $\Lambda(D)$ finite.

\section{Conclusion} \label{sec:mesh_conclusion}
%We have shown that the compress-forward strategy achieves the capacity of the SSMRSD mesh network asymptotically when the relay power grows. We note that when the relays can transmit at high power, they can communicate almost \emph{noiselessly} with each other and the destination. The best strategy for the nodes in this scenario is for them to cooperate to form a \emph{receive antenna array}~\cite{gastparkramer02}. This can be done via the compress-forward strategy.

The deployment of wireless networks will likely include mesh routers acting as relays.   For that reason, it makes sense to understand how these powerful relays should be used.  In this paper, we have taken a step in that direction using information theoretic ideas.  We have shown that the compress-forward strategy achieves the capacity of the SSMRSD mesh network asymptotically when the relays' powers are unconstrained. 

We note that when the relays can transmit at high power, they can communicate almost \emph{noiselessly} with each other and the destination.  A similar situation arises when the relays are clustered at the destination.  The best strategy (in an asymptotic sense) for the nodes in this scenario is for them to cooperate to form a \emph{receive antenna array}~\cite{gastparkramer02} and use compress-forward.  While the capacity achieving strategy is the same, we have observed that the convergence behaviors seem to be different.



% http://www.ctan.org/tex-archive/macros/latex/contrib/supported/IEEEtran/bibtex
%\bibliographystyle{IEEEtran.bst}

%---------- Bibliography -----------
\bibliography{bib}

\end{document}